\documentstyle[12pt]{article}        
\newlength{\dinwidth}
\newlength{\dinmargin}
\def\lsim{\mathrel{\rlap{\lower4pt\hbox{\hskip1pt$\sim$}}\raise1pt\hbox{$<$}}}
\def\gsim{\mathrel{\rlap{\lower4pt\hbox{\hskip1pt$\sim$}}
    \raise1pt\hbox{$>$}}}                
\begin{document}
\newcommand{\ds}{\displaystyle}
\newcommand{\la}[1]{\label{#1}}
\newcommand{\re}[1]{\ (\ref{#1})}
\newcommand{\nn}{\nonumber}
\newcommand{\be}{\begin{equation}}
\newcommand{\ee}{\end{equation}}
\newcommand{\ba}{\begin{eqnarray}}
\newcommand{\ea}{\end{eqnarray}}
\newcommand{\baz}{\begin{eqnarray*}}
\newcommand{\eaz}{\end{eqnarray*}}
\newcommand{\ct}[1]{${\cite{#1}}$}
\newcommand{\ctt}[2]{${\cite{#1}-\cite{#2}}$}
\newcommand{\bi}[1]{\bibitem{#1}}
\newcommand{\ep}{\epsilon}
\newcommand{\epm}{\varepsilon_{\mu\nu\lambda\sigma}}
\newcommand{\epa}{\varepsilon_{\mu\alpha\nu\beta}}
\newcommand{\epb}{\varepsilon_{\mu\nu\alpha\beta}}
\newcommand{\Sa}{S_{\mu\alpha\nu\beta}}
\newcommand{\gb}{\gamma_\beta}
\newcommand{\g}{\gamma_5}
\newcommand{\lc}{\lambda^c}
\newcommand{\bq}{\bar q}

\renewcommand{\arraystretch}{2}
\begin{titlepage}

\large
\thispagestyle{empty}
\normalsize
\begin{flushleft}
DESY 96--175\\
{\tt hep-ph/9612318}\\
December  1996
\end{flushleft}
\vspace*{2cm}
\begin{center}
\LARGE
{\bf On the Twist 2 and Twist 3 Contributions} \\

\vspace{1mm}
{\bf to the Spin-dependent Electroweak} \\

\vspace{1mm}
{\bf Structure Functions}

\vspace{2.0cm}
\large
Johannes Bl\"umlein$^a$
 and Nikolai Kochelev$^{a,b}$

\vspace{2.0cm}
\large {\it
 $^a$DESY--Zeuthen \\

\vspace{0.1cm}
Platanenallee 6, D--15735 Zeuthen, Germany }\\
\large {\it
 $^b$Bogoliubov Laboratory of Theoretical Physics, JINR,}\\

\vspace{0.1cm}
{\it
RU--141980 Dubna, Moscow Region, Russia }

\vspace{2cm}
\normalsize
{\bf Abstract} \\
\end{center}
\noindent
The twist 2 and twist 3 contributions of the polarized deep-inelastic
structure functions are calculated both for neutral and charged current 
interactions using the operator product expansion in lowest order in QCD.
The relations between the different structure functions are determined.
New integral relations are derived between the twist 2 contributions of
the structure functions $g_3(x,Q^2)$ and $g_5(x,Q^2)$ and between
combinations of the twist 3 contributions to the structure functions 
$g_2(x,Q^2)$ and $g_3(x,Q^2)$. The sum rules for polarized deep inelastic
scattering are discussed in detail.

\end{titlepage}
\newpage

\section{Introduction}
\label{sect1}
%
The investigation of the nucleon structure by polarized deep-inelastic
lepton scattering off polarized targets revealed a rich structure of
phenomena during the last years~\cite{REV}. So far only the case of
polarized deep-inelastic photon scattering has been studied experimentally
for lower
values of $Q^2$. In the future possible polarized deep-inelastic
scattering experiments at the RHIC collider and HERA may be considered
in a much wider kinematic range~\cite{JB95} in which the contributions
of also the weak currents are important. For this general case the
scattering cross sections are determined by five different
polarized structure functions,
if lepton mass effects are disregarded.

The light--cone expansion~\cite{LCO} proofed to be one of the most
powerful techniques to describe the behavior of deep-inelastic scattering
structure functions in the Bjorken limit~\cite{BjL}.
Whereas the case
of unpolarized deep-inelastic scattering and polarized scattering with
pure photon exchange are well understood~\cite{DIS} still
differing results are reported for polarized deep-inelastic scattering
including weak neutral and charged current interactions,
see e.g. refs.~\cite{a1}--\cite{a17}.

In previous investigations different techniques were used to derive
relations between the polarized structure functions.
In refs.~\cite{a1}--\cite{BAR} the structure functions were calculated in
the parton model. Some of the investigations deal with the
case of longitudinal polarization only~\cite{a3}.
In other studies light-cone current algebra~\cite{DIC}--\cite{A1C} and
the operator product expansion were
used~\cite{AR}--\cite{a77}. Furthermore, the structure
functions $g_1^{em}$
and $g_2^{em}$ were also
calculated in the covariant parton model~\cite{LP} in
refs.~\cite{a1B,a17}.
Still a thorough agreement between the different approaches has not
been obtained.

Unlike the structure functions in deep-inelastic scattering off
unpolarized targets, two of the structure  functions of polarized
nucleons, $g_2(x,Q^2)$ and $g_3(x,Q^2)$, contain  besides  the twist~2
terms  also twist~3 contributions.
The determination of these contributions to polarized structure
functions is a great challenge to experiment. It requires a precise
measurement
of the deep-inelastic scattering structure functions for transversely
polarized targets\footnote{For first experimental results
see ref.~\cite{TRAEX}.}. 
Polarized deep-inelastic scattering allows thus to test
predictions of QCD in a {\it new} domain.
Furthermore, to unfold also the flavor structure
of these contributions charged current polarized deep-inelastic
scattering has to be studied\footnote{This possibility has been
considered already earlier in ref.~\cite{BAR}.} which is experimentally
extremely difficult.

It is the aim of the present paper to derive the relations for the 
complete set of the polarized structure functions including weak 
interactions, which are not associated with terms in the scattering cross
section vanishing as $m_{lepton} \rightarrow 0$, accounting both for
their twist 2 and twist 3 contributions.
The calculation is performed applying the operator product expansion.

As it turns out, the twist 2 contributions for only two out of the five
polarized structure
functions, corresponding to the respective current combinations, are
linearly independent. Therefore three
linear operators have to exist which
determine the twist 2 contributions of the
remaining three structure
functions over a basis of two in lowest order QCD.
Two of them are given by the
Wandzura--Wilczek~\cite{WW} relation  and a
relation by Dicus~\cite{DIC}. A third new relation is derived.

The  structure function $g_3(x,Q^2)$
contributes only in the exchange of a weak current.
We derive also
a {\it new} relation between  combinations of
its twist 3 contribution and corresponding contributions
to the structure
function $g_2(x,Q^2)$.
New sum rules based on this  relation can be obtained. 
A brief summary of
our results for  the twist 2 terms for neutral current deep-inelastic
scattering
has already
appeared~\cite{BK}.

The paper is organized as follows. In section~2 basic notations are
introduced and the Born cross sections for polarized deep-inelastic
scattering are discussed. The structure of the forward Compton amplitude 
and the crossing relations for the case of neutral and charged
electroweak currents are derived in section~3. In section~4 a detailed
derivation of the operator product expansion is presented. Relations
between the moments of the structure functions are derived
in section~5. In section~6 a critical account is given on
sum rules discussed in the literature, and section~7 contains
the conclusions.


\section{The Born Cross Sections}
\label{sect2}

\vspace{1mm}
\noindent
The differential Born cross section for polarized lepton--polarized
nucleon scattering reads
\be
\label{e1}
\frac{d^3 \sigma}{dx dy d\theta} = \frac{y \alpha^2}
{Q^4} \sum_{i} \eta_i(Q^2) L^{\mu\nu}_i W_i^{\mu\nu}.
\ee
Here the index $i$ denotes the different current combinations,
i.e. $i = |\gamma|^2, |\gamma Z|, |Z|^2$ for the
neutral, and
$i = |W^-|^2$ or $|W^+|^2$  for
the charged current interactions. $\theta$ is the azimuthal angle
of the final-state lepton,
$x = Q^2/(2 P.q)$ and
$y =  P.q/k.P$ are the Bjorken variables, with $q = k - k'$ the
four momentum transferred, $k$ and $k'$ the initial and final-state
lepton, $P$ the  proton 4-momenta,
respectively, and $Q^2 = -q^2$.
The factors $\eta_i(Q^2)$ denote the ratios of the
corresponding propagator terms to the photon propagator squared,
\begin{eqnarray}
\label{etas}
\eta^{|\gamma|^2}(Q^2)  &=& 1 ,\nonumber\\
\eta^{|\gamma Z|}(Q^2)  &=& \frac{G_FM_Z^2}
{2\sqrt{2}\pi\alpha}\frac{Q^2}{Q^2+M_Z^2},
\nonumber\\
\eta^{|Z|^2}(Q^2)       &=& (\eta^{|\gamma Z|})^2(Q^2), \nonumber\\
\eta^{|W^{\pm}|^2}(Q^2) &=& \frac{1}{2}
\left( \frac{G_FM_W^2}{4\pi\alpha} \frac{Q^2}{Q^2+M_W^2} \right) ^2.
\end{eqnarray}
$\alpha$ denotes the fine structure constant, $G_F$ is the Fermi
constant and $M_Z$ and $M_W$ are the $Z$ and $W$~boson masses.

The leptonic tensor has the following form:
\be
\label{e2}
L_{\mu\nu}^i=\sum_{\lambda^\prime}\left[\bar
u(k^\prime,\lambda^\prime)\gamma_\mu(g_V^{i_1}+g_A^{i_1}
\gamma_5)u(k,\lambda)\right]^\ast\bar
u(k^\prime,\lambda^\prime)\gamma_\nu(g_V^{i_2}
+g_A^{i_2}\gamma_5)u(k,\lambda).
\ee
$\lambda$ and $\lambda'$  denote the helicity of the initial and
final-state lepton. The indices $i_1$ and $i_2$ refer to the currents
forming the combinations $i$ in eq.~(\ref{e1}).
The vector and axial vector couplings are
\begin{equation}
\label{ac}
\begin{array}{lcllcl}
g_V^{\gamma} &=& 1, & g_A^{\gamma} &=& 0, \\
g_V^Z&=&\ds
-\frac{1}{2}+2 \sin^2\theta_W,& g_A^Z &=&\ds \frac{1}{2}, \\
g_V^{W^-} &=& 1, &g_A^{W^-} &=& -1,
\end{array}
\end{equation}
\renewcommand{\arraystretch}{2}

\noindent
for 
negatively charged 
initial-state 
leptons and neutrinos.
Here $\theta_W$ denotes the weak mixing angle.
For
positively
charged leptons and antineutrinos the sign of the axial vector couplings
is reversed.

The hadronic tensor is given by
\ba
W_{\mu\nu}^{i}=\frac{1}{4\pi}\int d^4x e^{iqx}
\langle PS\mid[{J_\mu^{i_1}(x)}^\dagger ,J_\nu^{i_2}(0)]\mid PS\rangle.
\ea
$S$ denotes the nucleon spin 4-vector with,
$S\cdot P=0$. In the following we normalize $S^2 = -M^2$, where $M$
is the nucleon mass.
In the  framework of the quark--parton model the currents
$J_{\mu}^j$ read
\ba
\label{eqCUR1}
J_\mu^j(x)=\sum_{f,f'}\bar q_f'
(x)\gamma_\mu(g_V^{j,f}+g_A^{j,f}\gamma_5)
q_f(x)U_{ff^\prime},
\ea
where  $g_{V,A}^{j,f}$ are the electroweak
couplings of the quark labeled by $f$.
For charged current interactions
$U_{ff^\prime}$ denotes the
Cabibbo-Kobayashi-Maskawa matrix and $ g_V=1, {\ } g_A=-1$, whereas
for neutral current
interactions  $U_{ff^\prime} = \delta_{ff^\prime}$,
$ g_V^q = e_q, g_A = 0$ for $\gamma $, and
\be
\label{e111}
\begin{array}{lcrlcrl}
g_V^q &=& \ds
\frac{1}{2}-\frac{4}{3}\sin^2\theta_W,
& g_A^q &=& \ds
-\frac{1}{2}, &~~{\rm for}~~q=u,c \\
g_V^q &=&\ds
-\frac{1}{2}+\frac{2}{3}\sin^2\theta_W,
& g_A^q &=& \ds
\frac{1}{2}, &~~{\rm for}~~ q=d,s
\end{array}
\ee
\renewcommand{\arraystretch}{1.5}
for  $Z $ boson exchange. For  antiquarks the sign of  $g_A^q$
reverses.

The hadronic tensor is constructed
requiring
Lorentz and time-reversal invariance, as well as  current
conservation.\footnote{
Hadronic tensors with
a violation of
current conservation  have
been discussed in the literature, see e.g. ref.~\cite{a7,FRANKF}.}
The general structure of the hadronic tensor is
\begin{eqnarray}
\label{e6}
W_{\mu\nu}&=&(-g_{\mu\nu}+\frac{q_\mu q_\nu}{q^2})F_1(x,Q^2)+
\frac{\hat{P}_\mu\hat{P}_\nu}{P\cdot q}F_2(x,Q^2)-
 i\epm\frac{q^\lambda P^\sigma}{2P\cdot q}F_3(x,Q^2)\nn\\
& &~~+  i\epm\frac{q^\lambda S^\sigma}{P\cdot q}g_1(x,Q^2)+
i\epm\frac{q^{\lambda}(P\cdot qS^\sigma-S\cdot qP^\sigma)}
{(P\cdot q)^2}g_2(x,Q^2)\nn\\
& &~~+[\frac{\hat{P_\mu}\hat{S_\nu}+\hat{S_\mu}\hat{P_\nu}}{2}-
S\cdot q\frac{\hat{P_\mu}\hat{P_\nu}}{(P\cdot q)}]
\frac{g_3(x,Q^2)}{P\cdot q}\nn\\
& &~~+
S\cdot q\frac{\hat{P_\mu}\hat{P_\nu}}{(P\cdot q)^2}
g_4(x,Q^2)+
(-g_{\mu\nu}+\frac{q_\mu q_\nu}{q^2})\frac{(S\cdot q)}
{P\cdot q} g_5(x,Q^2),
\end{eqnarray}
with
\begin{eqnarray}
\hat{P_\mu}=P_\mu-\frac{P\cdot q}{q^2} q_{\mu}
,{\ }~~~~~~~~
\hat{S_\mu}=S_\mu-\frac{S\cdot q}{q^2} q_{\mu}.
\end{eqnarray}
Here the current indices were suppressed. The hadronic tensor depends
in general on three unpolarized structure functions $F_i$ and
five polarized structure functions $g_i$. The notation for the structure
functions  $F_i$, $g_1$ and $g_2$ is widely unique in the literature,
however, different notations are used for the structure functions $g_3,
g_4$ and $g_5$. For later comparisons, we list the main conventions
used by other authors in Table~1.

From eqs.~\re{e1},\re{e2} and \re{e6} one  obtains the differential
scattering cross sections of
a lepton with helicity $\lambda$ off a polarized nucleon.
For convenience we will consider two projections of the nucleon spin
vector, choosing
the spin direction longitudinally and transversely to the nucleon
momentum. In the nucleon rest frame one has
\begin{eqnarray}
S_L &=& (0, 0, 0, M), \nonumber\\
S_T &=& M (0, \cos\alpha, \sin\alpha,0).
\end{eqnarray}

\vspace{7mm}
\begin{center}
\begin{tabular}{||c||c|c|c|c|c||}\hline\hline
{\sf  our  notation}      & {\sf \cite{a1}} &
{\sf \cite{BAR}}& {\sf \cite{a3}} &
{\sf \cite{DIC}} &  {\sf \cite{WRAY}}\\ \hline \hline
$g_1$ & $g_1$ &  $g_1$      & $g_1$     & $F_3+F_4$ & $2(F_1+F_2) $\\
$g_2$ & $g_2$ &  $g_2$      & $g_2$  & $-F_4$ & $-2F_2 $\\
$g_3$ & $-g_3$ &  $g_5$      & $(g_4-g_5)/2$  &
$2F_8$ & $4F_4 $ \\
$g_4 $
& $g_4-g_3$
& $g_4+g_5$
&$g_4$ &$2F_8+F_7$ & $2(F_6+2F_4)$ \\
$g_5$
& $-g_5$
&$ -g_3$
&$g_3$ & $F_7/(2x)-F_6$ &$F_7/(2x)-2F_5$
\\  \hline \hline
      & {\sf \cite{A1C}} &
{\sf \cite{a7, FRANKF}}& {\sf \cite{a77}} &
{\sf \cite{a1B}} &  {\sf \cite{BC}}\\ \hline \hline
$g_1$ & $G_1+G_2$ &  $g_1$      & $g_1$     & $\tilde{F_1}$ &
$M\nu {\rm Im}~G_1/\pi $\\
$g_2$ & $-G_2$ &  $g_2$   
   & $g_2$  & $\tilde{F_2}/(2x)$ & $-\nu^2 {\rm Im}~G_2/(M\pi) $\\
$g_3$ & $2G_3$ &  $b_1+b_2$      & $(A_2-A_3)/2$  &
$-2\tilde{F_3}$ & $ $ \\
$g_4 $
& $2G_4+G_5$
& $a_2+b_1+b_2 $
&$A_2$ 
&$$ 
& $$ \\
$g_5$
& $-G_4$
&$a_1$
&$A_1$
& $$
&$$

\\  \hline \hline
\end{tabular}
\end{center}

\vspace{2mm}
\noindent
{\sf Table~1: A comparison of different conventions to denote the
polarized nucleon structure functions.}

\vspace{5mm}

\vspace{5mm}

The scattering cross section for longitudinal nucleon polarization reads
\ba
\frac{d^2\sigma(\lambda, \pm S_L)}{dxdy}
&=& 2 \pi S  \frac{\alpha^2}{Q^4}
\sum_i C_i \eta_i(Q^2) \left \{ y^2 2xF_1^i + 2\left (1 - y
- \frac{x y M^2}{S} \right) F_2^i
- 2\lambda y \left( 1-\frac{y}{2} \right ) xF_3^i \right. \nn\\
&\pm& \left [
-2\lambda y \left( 2-y-\frac{2 x y M^2}{S} \right) xg_1^i +
8 \lambda \frac{y x^2 M^2}{S} g_2^i
 + \frac{4 x M^2}{S} \left ( 1 - y - \frac{x y M^2}{S} \right)
 g_3^i \right.
\nn\\
&-& 2  \left. \left.
\left (1 + \frac{2 x M^2}{S} \right)
\left (1 - y - \frac{x y M^2}{S} \right ) g_4
- 2 x y^2 \left (1 + \frac{2 x M^2}{S} \right) g_5^i \right] \right
\}.
\ea
Correspondingly, for transversely polarized nucleons one obtains
\newpage
\ba
\frac{d^3\sigma(\lambda, \pm S_T)} {dx dy d\phi} &=&
S \frac{\alpha^2}{Q^4} \sum_i
C_i \eta_i(Q^2)
\left \{y^2 2 x F_1^i + 2 \left (1 - y - \frac{x y M^2}{S}
\right ) F_2^i - 2 \lambda y \left (1 - \frac{y}{2} \right) xF_3^i
\right. \nn\\
&\pm&
 2 \sqrt{\frac{M^2}{S}} \sqrt{x y \left [1 - y - \frac{x y M^2}{S}
\right]} \cos(\alpha-\phi) \left [
-2\lambda y x g_1^i -
4\lambda  g_2^i \right.
\nn\\
&-& \left. \left. \frac{1}{y} \left (2 - y
- \frac{2 x y M^2}{S} \right) g_3^i
+ \frac{2}{y}  \left (1 - y - \frac{x y M^2}{S} \right) g_4
+  y^2 2 x g_5^i \right] \right \}.
\ea
Here
$C^\gamma = 1$, $C^{\gamma Z} = g_V + \lambda g_A$,
$C^Z = (g_V + \lambda g_A)^2$, and $C^{W^{\pm}} = (1 \pm \lambda)$.
As well known, the contributions of the structure functions $g_2^i$ and
$g_3^i$ are suppressed by a factor of $M^2/S$
in the longitudinal spin asymmetries
$\Delta^L = d^2 \sigma(\lambda, S_L) - d^2 \sigma(\lambda, -S_L)$.
However, they contribute at the same strength as the other structure
functions to the transverse spin asymmetries
$\Delta^T = d^2 \sigma(\lambda, S_T) - d^2 \sigma(\lambda, -S_T)$.

\section{The Forward Compton Amplitude}
\label{sect3}

\vspace{1mm}
\noindent
The forward Compton amplitude $T_{\mu\nu}$
is related to the hadronic tensor by
\ba
\label{eqWT}
W_{\mu\nu}^{i}=\frac{1}{2\pi} {\rm Im}{\ }T_{\mu\nu}^{i},
\ea
with
\ba
\label{eqTT}
T_{\mu\nu}^{i} = i \int d^4xe^{iqx}
\langle PS\mid (T{J_\mu^{i_1}}^\dagger (x)J_\nu^{i_2}(0)) |PS\rangle .
\ea
It can be  represented in terms of the amplitudes $T^i_k$ and
$A^i_k$ as
\begin{eqnarray}
T_{\mu\nu}^i
&=&
(-g_{\mu\nu}+\frac{q_\mu q_\nu}{q^2})T_1^i(q^2,\nu)+
\frac{\hat{P}_\mu\hat{P}_\nu}{M^2}T_2^i(q^2,\nu)-
i\epm\frac{q_\lambda P\sigma}{2M^2}T_3^i(q^2,\nu)\nn\\
& &~~+
 i\epm\frac{q^\lambda S^\sigma}{M^2}A_1^i(q^2,\nu)+
i\epm\frac{q^{\lambda}(P\cdot qS^\sigma-S\cdot q
P^\sigma)}{M^4}A_2^i(q^2,\nu)\nn\\
& &~~+
[\frac{\hat{P_\mu}\hat{S_\nu}+\hat{S_\mu}\hat{P_\nu}}{2}-
S\cdot q\frac{\hat{P_\mu}\hat{P_\nu}}{(P\cdot q)}]
\frac{A_3^i(q^2,\nu)}{M^2}\nn\\
& &~~+
S\cdot q\frac{\hat{P_\mu}\hat{P_\nu}}{M^4}A_4^i(q^2,\nu)+
(-g_{\mu\nu}+\frac{q_\mu q_\nu}{q^2})\frac{S\cdot q}{M^2}A_5^i(q^2,\nu),
\label{e7}
\end{eqnarray}
where $\nu = P.q$.
The structure functions $g_i(x,Q^2)$ and amplitudes $A_i(q^2,\nu)$ are
related by
\ba
g_{1,3,5}(x,Q^2)&=&\frac{1}{2 \pi}
\frac{\nu}{M^2}{\rm Im} A_{1,3,5}(q^2,\nu), \nonumber\\
g_{2,4}(x,Q^2)&=&\frac{1}{2 \pi}\frac{\nu^2}{M^4}{\rm Im} 
A_{2,4}(q^2,\nu).
\label{rel}
\ea
Subsequently we will consider the polarized part of $T^i_{\mu \nu}$ only.

For neutral current interactions the current operators obey
\be
{J_\mu^{\gamma, Z}}^\dagger= J_\mu^{\gamma, Z}.
\label{neut}
\ee
Therefore, the crossing relation for the amplitude for $q \rightarrow -q,
P \rightarrow P$
reads
\be
\label{eqCR1}
T_{\mu\nu}^{i}(q^2, -\nu) = T_{\nu\mu}^{i}(q^2, \nu).
\ee
The corresponding relations
for the amplitudes $A_i^{\rm NC}(q^2, \nu)$ are
\begin{eqnarray}
A_{1,3}^{\rm NC}(q^2, -\nu) &=& ~A_{1,3}^{\rm NC}(q^2, \nu), \nonumber\\
A_{2,4,5}^{\rm NC}(q^2, -\nu) &=& -A_{2,4,5}^{\rm NC}(q^2, \nu).
\end{eqnarray}
Furthermore,
the amplitudes obey the following
forward dispersion relations~:
\begin{eqnarray}
A_{1,3}^{\rm NC}(q^2, \nu) &=& 
\frac{2}{\pi} \int_{Q^2/2}^{\infty} d\nu' \frac{\nu'} {\nu'^2 - \nu^2}
 {\rm Im}~A_{1,3}^{\rm NC}(q^2, \nu^\prime), \nn\\
A_{2,4,5}^{\rm NC}(q^2, \nu) &=& \frac{2}{\pi}
\int_{Q^2/2}^{\infty} d \nu' \frac{\nu}
{\nu'^2 - \nu^2} {\rm Im}~A_{2,4,5}^{\rm NC}(q^2, \nu').
\label{ndr}
\end{eqnarray}

For the charged current interactions it is suitable to study the
linear combination of amplitudes
\ba
\label{eqCOM}
T_{\mu\nu}^{\pm}(q^2, \nu) = T_{\mu\nu}^{W^-}(q^2, \nu)
                        \pm T_{\mu\nu}^{W^+}(q^2, \nu).
\ea
Due to the transformation
\be
{J_\mu^{W^\pm}}^\dagger= J_\mu^{W^\mp},
\label{char}
\ee
the following crossing relations hold~:
\be
T^{\pm}(q^2, -\nu) = \pm T^{\pm}(q^2, \nu).
\ee
Correspondingly, one obtains for the combination of the
amplitudes
\be
A_i^{\pm}={A_i}^{W^-}\pm {A_i}^{W^+}
\ee
the relations
\ba
A^\pm_{1,3}(q^2,-\nu) &=&
\pm A^\pm_{1,3}(q^2,\nu),{\ }{\ }\nn \\
A^\pm_{2,4,5}(q^2,-\nu) &=&
\mp A^\pm_{2,4,5}(q^2,\nu).
\label{cr}
\ea
The respective dispersion relations for $A_i^+(q^2,\nu)$ and 
$A_i^-(q^2,\nu)$ are:
\begin{eqnarray}
A_{1,3}^+(q^2, \nu) &=& \frac{2}{\pi} \int_{Q^2/2}^{\infty} d
\nu' \frac{\nu'} {\nu'^2 - \nu^2} {\rm Im} A_{1,3}^+(q^2,
\nu^\prime),\nn \\
A_{2,4,5}^+(q^2, \nu)&=& \frac{2}{\pi}
\int_{Q^2/2}^{\infty} d \nu' \frac{\nu}
{\nu'^2 - \nu^2} {\rm Im} A_{2,4,5}^+(q^2, \nu'),\nn\\
A_{1,3}^-(q^2, \nu) &=& \frac{2}{\pi} \int_{Q^2/2}^{\infty} d
\nu' \frac{\nu'} {\nu(\nu'^2 - \nu^2)} {\rm Im}\left[
 \nu^\prime A_{1,3}^-(q^2,
\nu^\prime) \right], \nn \\
A_{2,4}^-(q^2, \nu) &=& \frac{2}{\pi} \int_{Q^2/2}^{\infty} d
\nu' \frac{\nu'} {\nu^2(\nu'^2 - \nu^2)} {\rm Im}\left[
{\nu^\prime}^2  A_{2,4}^-(q^2, \nu^\prime) \right],
\nn\\
A_{5}^-(q^2, \nu) &=& \frac{2}{\pi}
\int_{Q^2/2}^{\infty} d \nu' \frac{\nu^\prime}
{\nu'^2 - \nu^2} {\rm Im}~A_{5}^-(q^2, \nu').
\label{cdr2}
\end{eqnarray}

For the case of charged current interactions we introduce the structure
function combinations
\be
g_i^{\pm}(x,Q^2) = g_i^{W^-}(x,Q^2) \pm g_i^{W^+}(x,Q^2).
\label{eqGIPM}
\ee
The integral representations of the amplitudes $A_i^{\rm NC}$ and
$A_i^{\pm}$ can be finally expressed by the moments of the corresponding
structure functions as
\ba
A_{1,3}^{NC,+}(q^2,\nu)&=&\frac{4M^2}{\nu}
\sum_{n=0,2...}\frac{1}{x^{n+1}}
\int_0^1dyy^ng_{1,3}^{NC,+}(y,Q^2),\nn\\
A_{2,4}^{NC,+}(q^2,\nu)&=&\frac{4M^4}{\nu^2}
\sum_{n=2,4...}\frac{1}{x^{n+1}}
\int_0^1dyy^ng_{2,4}^{NC,+}(y,Q^2),\nn\\
A_5^{NC,+}(q^2,\nu)&=&\frac{4M^2}{\nu}
\sum_{n=1,3...}\frac{1}{x^{n+1}}\int_0^1dyy^ng_{5}^{NC,+}(y,Q^2),
\label{tayNC}
\ea
and
\ba
A_{1,3}^{-}(q^2,\nu)&=&\frac{4M^2}{\nu}
\sum_{n=-1,1...}\frac{1}{x^{n+1}}
\int_0^1dyy^ng_{1,3}^{-}(y,Q^2),\nn\\
A_{2,4}^{-}(q^2,\nu)&=&\frac{4M^4}{\nu^2}
\sum_{n=-1,1...}\frac{1}{x^{n+1}}
\int_0^1dyy^ng_{2,4}^{-}(y,Q^2),\nn\\
A_{5}^{-}(q^2,\nu)&=&\frac{4M^2}{\nu}
\sum_{n=0,2...}\frac{1}{x^{n+1}}
\int_0^1dyy^ng_{5}^{-}(y,Q^2),
\label{tay-}
\ea
performing a Taylor expansion of eqs.~(\ref{ndr}, \ref{cdr2}) and using
eqs.~(\ref{rel}, \ref{eqGIPM}), where $x = Q^2/(2 \nu)$ and
$y = Q^2/(2 \nu')$.

\section{ Operator product expansion}
\label{sect4}

\vspace{1mm}
\noindent
The operator product expansion is one of the most general formalisms
to analyze the properties of the structure functions in deep-inelastic
scattering. 
We  apply it to
the $T$-product of two electroweak currents,
\ba
\hat{T}^i_{\mu \nu} = T({J_\mu^{i_1}}^\dagger (x)J_\nu^{i_2}(0)).
\ea
Near the light cone one obtains for neutral currents
\ba
\label{T1}
\hat{T}^{NC}_{\mu \nu} &=& \bar q(x) \gamma_{\mu}
(g_{V_1} + g_{A_1} \gamma_5) S(~~x) \gamma_{\nu}
(g_{V_2} + g_{A_2}\gamma_5)P^+ q(0)
  \nn\\
&{+}& \bar q(0) \gamma_{\nu} (g_{V_2}+g_{A_2}\gamma_5) S(-x)
\gamma_{\mu} (g_{V_1} + g_{A_1} \gamma_5) P^+ q(x),
\ea
and for the charged current combinations
\be
\hat{T}^{\pm}_{\mu \nu}=\hat{T}^{W^-}_{\mu \nu}\pm 
\hat{T}^{W^+}_{\mu \nu},
\ee
\ba
\label{T2}
\hat{T}^{\pm}_{\mu \nu}
 &=& \bar q(x) \gamma_{\mu}
(g_{V_1} + g_{A_1} \gamma_5) S(~~x) \gamma_{\nu}
(g_{V_2} + g_{A_2}\gamma_5)P^\pm q(0)
  \nn\\
&{\pm}& \bar q(0) \gamma_{\nu} (g_{V_2}+g_{A_2}\gamma_5) S(-x)
\gamma_{\mu} (g_{V_1} + g_{A_1} \gamma_5) P^\pm q(x).
\ea
Here we used
the projectors
\be
P^+= {\bf 1} ,~~~~~~~~~~~~~~~~~~~~~~~~~~~~~P^-= {\bf \tau_3} ,
\ee 
with
\ba
{\bf \tau_3} = \left ( \begin{array}{cc} 1 & 0 \\
0 & -1 \end{array} \right ),
\nn
\ea
and suppressed the flavor indices and the Cabibbo-Kobayashi-Maskawa
matrix  for brevity.
If not stated otherwise,
we will not distinguish the neutral current case from the
charged current + combination subsequently. The quark and antiquark
states in eqs.~(\ref{T1}, \ref{T2}) are  understood as singlets
for neutral current interactions and as
doublets for charged current
interactions.
In eqs.~\re{T1} and \re{T2}
\ba
S(x) \approx \frac{2i \not \!{x}}{(2 \pi)^2(x^2 - i0)^2}
\label{prop}
\ea
denotes the
free quark propagator.

We rewrite the above relations for $\hat{T}^{+}_{\mu \nu}$
and $\hat{T}^{-}_{\mu \nu}$ using
\begin{eqnarray}
\gamma_\mu \not \!{x} \gamma_\nu &=& x^\alpha [\Sa\gb-i\epa\gb\g],
\nonumber\\
\Sa &=& g_{\mu \alpha} g_{\nu \beta} + g_{\mu \beta} g_{\nu \alpha}
- g_{\mu \nu} g_{\alpha \beta}.
\label{e9}
\end{eqnarray}
For the spin
dependent part of $\hat{T}^i_{\mu \nu}$ one obtains
\ba
\label{eqT2}
\hat{T}^{+}_{\mu \nu, spin}  =  \frac{2ix^\alpha}{(2\pi)^2(x^2-i0)^2}
\left \{ -i (g_{V_1} g_{V_2} + g_{A_1} g_{A_2}) \epa
u^{\beta}_+
+ (g_{V_1} g_{A_1} + g_{A_1} g_{V_2}) \Sa
u^{\beta}_- \right \},
\ea
\ba
\label{eqT-}
\hat{T}^{-}_{\mu \nu, spin}  =  \frac{2ix^\alpha}{(2\pi)^2(x^2-i0)^2}
\left \{ -i (g_{V_1} g_{V_2} + g_{A_1} g_{A_2}) \epa
v^{\beta}_-
+ (g_{V_1} g_{A_1} + g_{A_1} g_{V_2}) \Sa
v^{\beta}_+ \right \},
\ea
with
\ba
u^{\beta}_{\pm}   =
\bq(x) \gamma^{\beta} \g P^+ q(0)
\pm
\bq(0) \gamma^{\beta} \g P^+ q(x),
\ea
\ba
v^{\beta}_{\pm}   =
\bq(x) \gamma^{\beta} \g P^- q(0)
\pm
\bq(0) \gamma^{\beta} \g P^- q(x).
\ea
The operators in eq.~(\ref{eqT2}) can be represented by the following
Taylor series around $x=0$
\ba
\bq(x) \gamma^{\beta} \g P^\pm q(0) &=& \sum_n
\frac{(-1)^n}{n!}x_{\mu_1}...x_{\mu_n} \bq(0) \gamma^{\beta} \g
D^{\mu_1}... D^{\mu_n}  P^\pm q(0), \nn\\
\bq(0) \gamma^{\beta} \g P^\pm q(x) &=& \sum_n
\frac{(+1)^n}{n!}x_{\mu_1}...x_{\mu_n} \bq(0) \gamma^{\beta} \g
D^{\mu_1}... D^{\mu_n}P^\pm q(0),
\ea
for which
\ba
\label{eqT3}
\hat{T}^{+}_{\mu \nu,spin} &=&
\frac{4 i x^\alpha}{(2 \pi)^2 (x^2-i0)^2}
\left \{ -i (g_{V_1} g_{V_2} + g_{A_1} g_{A_2}) \epa
 \rho^{+\beta}_{+} \right. \nonumber\\
& &\left.~~~~~~~~~~~~~~~~~~~~~-~(g_{V_1} g_{A_2}
+ g_{A_1} g_{V_2}) \Sa \rho^{+\beta}_{-} \right \},
\ea
\ba
\label{eqT3-}
\hat{T}^{-}_{\mu \nu, spin} &=& 
\frac{4 i x^\alpha}{(2 \pi)^2 (x^2-i0)^2}
\left \{ -i (g_{V_1} g_{V_2} + g_{A_1} g_{A_2}) \epa
 \rho^{-\beta}_{-} \right. \nonumber\\
& &\left.~~~~~~~~~~~~~~~~~~~~~-~(g_{V_1} g_{A_2}
+ g_{A_1} g_{V_2}) \Sa \rho^{-\beta}_{+} \right \}
\ea
is obtained. Here we used the abbreviations
\ba
\rho^{\pm\beta}_+ &=&
 \sum_{n~{\rm even}}
\frac{1}{n!} x_{\mu_1}...x_{\mu_n} \bq(0) \gamma^{\beta} \g
D^{\mu_1}... D^{\mu_n}P^\pm q(0),
\nonumber\\
\rho^{\pm\beta}_{-} &=&
 \sum_{n~{\rm odd }}
\frac{1}{n!} x_{\mu_1}...x_{\mu_n} \bq(0) \gamma^{\beta} \g
D^{\mu_1}... D^{\mu_n}P^\pm q(0)\nn,
\ea
for which
it is convenient to define (cf.~e.g.~\cite{RLJ})
\ba
\rho^{\pm\beta}_{\pm} \equiv  \sum_{n~{\rm even}/{\rm odd}}
\frac{1}{n!} i x_{\mu_1}...i x_{\mu_n}
\Theta^{\pm\beta\{\mu_1 ... \mu_n\}}
\ea
for later analysis.
The Fourier transforms of eqs.~(\ref{eqT3}, \ref{eqT3-}) read
\ba
\widehat{T}_{\mu \nu, spin}^{+}
&=&  i \int d^4x e^{iqx} \hat{T}_{\mu \nu, spin}^{+}
\nonumber\\
&=& \frac{1}{\pi^2} \int d^4x e^{iqx} \frac{x^{\alpha}}{(x^2 - i0)}
\left \{ i (g_{V_1} g_{V_2} + g_{A_1} g_{A_2}) \epa
 \rho^{+\beta}_{+} + (g_{V_1} g_{A_2}
+ g_{A_1} g_{V_2}) \Sa \rho^{+\beta}_{-} \right \}
\nonumber\\
&=& 
\left. -i (g_{V_1} g_{V_2} + g_{A_1} g_{A_2}) \epa  q^{\alpha}
\sum_{n~{\rm even}} q^{\mu_1} \ldots q^{\mu_n} \left ( \frac{2}{Q^2}
\right )^{n+1} \Theta^{+\beta\{\mu_1 \ldots \mu_n\}} \right.
\nonumber\\
&+&  \left.
(g_{V_1} g_{A_2} + g_{A_1} g_{V_2})[-g_{\mu\nu}\sum_{n~{\rm even}}
 q^{\mu_1} \ldots q^{\mu_n} \left ( \frac{2}{Q^2}
\right )^{n} \Theta^{+\mu_1\{\mu_2 \ldots \mu_n\}}\right.  
\nonumber\\
&+&  \left.
\sum_{n~{\rm even}}
 q^{\mu_1} \ldots q^{\mu_n} \left ( \frac{2}{Q^2}
\right )^{n+1} (
\Theta^{+\mu\{\nu \mu_1 \ldots \mu_n\}}+
\Theta^{+\nu\{\mu \mu_1 \ldots \mu_n\}})]  \right. ,
\label{ft1}
\ea
and
\ba
\widehat{T}_{\mu \nu, spin}^{-}
&=&  i \int d^4x e^{iqx} \hat{T}_{\mu \nu, spin}^{-}
\nonumber\\
&=& \frac{1}{\pi^2} \int d^4x e^{iqx} \frac{x^{\alpha}}{(x^2 - i0)}
\left \{ i (g_{V_1} g_{V_2} + g_{A_1} g_{A_2}) \epa
 \rho^{-\beta}_{-} + (g_{V_1} g_{A_2}
+ g_{A_1} g_{V_2}) \Sa \rho^{-\beta}_{+} \right \}
\nonumber\\
&=& 
\left. -i (g_{V_1} g_{V_2} + g_{A_1} g_{A_2}) \epa  q^{\alpha}
\sum_{n~{\rm odd}} q^{\mu_1} \ldots q^{\mu_n} \left ( \frac{2}{Q^2}
\right )^{n+1} \Theta^{-\beta\{\mu_1 \ldots \mu_n\}} \right.
\nonumber\\
&+&  \left.
(g_{V_1} g_{A_2} + g_{A_1} g_{V_2})[-g_{\mu\nu}\sum_{n~{\rm odd}}
 q^{\mu_1} \ldots q^{\mu_n} \left ( \frac{2}{Q^2}
\right )^{n} \Theta^{-\mu_1\{\mu_2 \ldots \mu_n\}}\right.  
\nonumber\\
&+&  \left.
\sum_{n~{\rm odd}}
 q^{\mu_1} \ldots q^{\mu_n} \left ( \frac{2}{Q^2}
\right )^{n+1} (
\Theta^{-\mu\{\nu \mu_1 \ldots \mu_n\}}+
\Theta^{-
\nu\{\mu \mu_1 \ldots \mu_n\}})]  \right. .
\label{ft2}
\ea

The operators $\Theta^{\pm\beta\{\mu_1 ... \mu_n\}}$ can be decomposed 
into
a symmetric and a remainder part, $\Theta_S$ and $\Theta_R$,
respectively
\ba
\Theta^{\pm\beta\{\mu_1 ... \mu_n\}}  =
\Theta_S^{\pm\beta\{\mu_1 ... \mu_n\}} +
\Theta_R^{\pm\beta\{\mu_1] ... \mu_n\}},
\ea
where
\ba
\Theta_S^{\pm\beta\{\mu_1 ... \mu_n\}}  &=&
\frac{1}{n+1} \left [
\Theta^{\pm\beta\{\mu_1 ... \mu_n\}}  +
\Theta^{\pm\mu_1 \{\beta ... \mu_n\}}  + ... +
\Theta^{\pm\mu_n \{\mu_1 ... \beta\}} \right], \\
\Theta_R^{\pm\beta\{\mu_1] ... \mu_n\}}  &=&
\frac{1}{n+1} \left [
\Theta^{\pm\beta\{\mu_1 ... \mu_n\}}  -
\Theta^{\pm\mu_1 \{\beta ... \mu_n\}}  +
\Theta^{\pm\beta \{\mu_1 \mu_2 ... \mu_n\}}  -
\Theta^{\pm\mu_2 \{\mu_1 \beta ... \mu_n\}} + ...~\right]. \nonumber
\ea
The nucleon matrix elements of these operators are
\ba
\label{MS}
\langle PS | \Theta_S^{\pm\beta\{\mu_1 ... \mu_n\}} |PS\rangle &=&
\frac{a_n^\pm}{n+1} \left [ S^{\beta} P^{\mu_1} P^{\mu_2} ... P^{\mu_n}
                      + S^{\mu1} P^{\beta} P^{\mu_2} ... P^{\mu_n}
                      + ... - {\sf traces} \right ], \\
\label{MR}
\langle PS | \Theta_R^{\pm\beta\{\mu_1] ... \mu_n\}} |PS\rangle &=&
\frac{d_n^\pm}{n+1} 
\left [ \left( S^{\beta} P^{\mu_1} - S^{\mu_1} P^{\beta}
\right ) P^{\mu_2} ... P^{\mu_n}  \right.
\nonumber\\
& &~~~~~~+
\left( S^{\beta} P^{\mu_2} - S^{\mu_2} P^{\beta}
\right ) P^{\mu_1} P^{\mu_3} ... P^{\mu_n}  \nonumber\\
& &~~~~~~+ \left.
\ldots + \left( S^{\beta} P^{\mu_n} - S^{\mu_n} P^{\beta}
\right ) P^{\mu_1} P^{\mu_3} ... P^{\mu_{n-1}} - {\sf traces} \right ].
\ea
The matrix elements $a_n^{\pm}$ and $d_n^{\pm}$ are the expectation
values of twist 2 and twist 3 operators, respectively.
For the contractions related to the structure functions $g_2^i$ one may
use a {\it simplified} version of (\ref{MR}), cf.~\cite{AR},
\ba
\label{MRsim}
\langle PS | \Theta_R^{\pm\beta\{\mu_1] ... \mu_n\}} |PS\rangle &\approx&
\frac{d_n^\pm}{n+1} \left [
\left( S^{\beta} P^{\mu_1} - S^{\mu_1} P^{\beta}
\right ) P^{\mu_2} ... P^{\mu_n}
-
 {\sf traces} \right ].
\ea
Note, however, that
this is {\it not} possible for the contractions emerging in the
case of the structure functions $g_3^i$.

Finally we obtain for the expectation values of the forward Compton
amplitude between nucleon states
\ba
{T}_{\mu \nu, spin}^{+}
&=& 
\left. -i (g_{V_1} g_{V_2} + g_{A_1} g_{A_2}) 
\frac{\epa  q^{\alpha}}{\nu}
\sum_{n~{\rm even}} 
 \frac{1}{x^{n+1}}
\Biggl [
\frac{a_n^++nd_n^+}{n+1}S^\beta+\frac{n(a_n^+-d_n^+)}{n+1}
\frac{(S.q)}{\nu}
P^\beta \Biggr ]
 \right.
\nonumber\\
&+&  \left.
(g_{V_1} g_{A_2} + g_{A_1} g_{V_2}) \Biggl \{
-g_{\mu\nu}\frac{(S.q)}{\nu}
\sum_{n~{\rm odd}}
 \frac{a_n^+}{x^{n+1}}
 \right.  
\nonumber\\
&+&  \left.
\frac{2}{\nu}\sum_{n~{\rm odd}}\frac{1}{x^n}
\Biggl [
\frac{2a_n^++(n-1)d_n^+}{n+1}
\Biggl
(\frac{S^\mu P^\nu+P^\mu S^\nu}{2}-\frac{P^\mu P^\nu}{\nu}(S.q) \Biggr )
   \right.
\nn\\ 
&+&  \left.
a_n^+\frac{P^\mu P^\nu}{\nu}(S.q) \Biggr ] \Biggr \},
   \right.
\label{ft3}
\ea
\ba
{T}_{\mu \nu, spin}^{-}
&=& 
\left. -i (g_{V_1} g_{V_2} + g_{A_1} g_{A_2}) \frac{\epa  q^{\alpha}}{\nu}
\sum_{n~{\rm odd}} 
 \frac{1}{x^{n+1}} \Biggl [
\frac{a_n^-+nd_n^-}{n+1}S^\beta+\frac{n(a_n^--d_n^-)}{n+1}
\frac{(S.q)}{\nu}
P^\beta \Biggr ]
 \right.
\nonumber\\
&+&  \left.
(g_{V_1} g_{A_2} + g_{A_1} g_{V_2}) \Biggl \{
-g_{\mu\nu}\frac{(S.q)}{\nu}
\sum_{n~{\rm even}}
 \frac{a_n^-}{x^{n+1}}
 \right.  
\nonumber\\
&+&  \left.
\frac{2}{\nu}\sum_{n~{\rm even}}\frac{1}{x^n} \Biggl [
\frac{2a_n^-+(n-1)d_n^-}{n+1}
\Biggl
(\frac{S^\mu P^\nu+P^\mu S^\nu}{2}-\frac{P^\mu P^\nu}{\nu}(S.q) \Biggr)
   \right.
\nn\\ 
&+&  \left.
a_n^-\frac{P^\mu P^\nu}{\nu}(S.q) \Biggr ]
   \right. \Biggr \}.
\label{ft4}
\ea
Here we
arranged  the structure as in eq.~(\ref{e7}) according to the 
contributions to the different amplitudes.

\newpage
\section{Relations between the Moments of Structure Functions}
\label{sect5}

\vspace{1mm}
\noindent
From eqs.~(\ref{ft3}) and (\ref{ft4}) one derives the following
representations for the amplitudes $A^{\rm NC}_i(q^2, \nu)$ and
$A^{\pm}_i(q^2, \nu)$~:
\ba
A_1^{NC,+}(q^2,\nu)&=&\frac{M^2}{\nu}
\sum_{\rm
n~even}\frac{((g_V^q)^ 2+(g_A^q)^2)a_n^{+q}}{x^{n+1}}\nn\\
A_2^{NC,+}(q^2,\nu)&=&\frac{M^4}{\nu^2}
\sum_{\rm
n~even}\frac{((g_V^q)^2+(g_A^q)^2)n(d_n^{+q}-a_n^{+q})}{x^{n+1}(n+1)}
\nn\\
A_3^{NC,+}(q^2,\nu)&=&\frac{M^2}{\nu}
\sum_{\rm
n~odd}\frac{4
g_V^qg_A^q(2a_n^{+q}+(n-1)d_n^{+q})}
{x^n(n+1)}
\nn\\
A_4^{NC,+}(q^2,\nu)&=&\frac{M^4}{\nu^2}
\sum_{\rm
n~odd}\frac{4g_V^qg_A^qa_n^{+q}}{x^n}
\nn\\
A_5^{NC,+}(q^2,\nu)&=&\frac{M^2}{\nu}
\sum_{\rm
n~odd}\frac{2g_V^qg_A^qa_n^{+q}}{x^{n+1}},
\label{rNC}
\ea
and
\ba
A_1^{-}(q^2,\nu)&=&\frac{M^2}{\nu}
\sum_{\rm
n~odd}\frac{((g_V^q)^ 2+(g_A^q)^2)a_n^{-q}}{x^{n+1}}\nn\\
A_2^{-}(q^2,\nu)&=&\frac{M^4}{\nu^2}
\sum_{\rm
n~odd}
\frac{((g_V^q)^2+(g_A^q)^2)n(d_n^{-q}-a_n^{-q})}{x^{n+1}(n+1)}
\nn\\
A_3^{-}(q^2,\nu)&=&\frac{M^2}{\nu}
\sum_{\rm
n~even}\frac{4g_V^qg_A^q(2a_n^{-q}+(n-1)d_n^{-q})}{x^n(n+1)}
\nn\\
A_4^{-}(q^2,\nu)&=&\frac{M^4}{\nu^2}
\sum_{\rm
n~even}\frac{
4g_V^qg_A^qa_n^{-q}}{x^n}
\nn\\
A_5^{-}(q^2,\nu)&=&\frac{M^2}{\nu}
\sum_{\rm
n~even}\frac{2g_V^qg_A^qa_n^{-q}}{x^{n+1}}.
\label{rC}
\ea
On the other hand, the representations eq.~(\ref{tayNC}) and (\ref{tay-})
are valid, from which the following
relations between the operator matrix
elements $a_n^{\pm,q}$ and $d_n^{\pm,q}$ and the moments of the structure
functions $g_i^{{\rm NC}, \pm}(x,Q^2)$ are obtained\footnote{Note
that
a misprint in  eq.~(17), \cite{BK},
was corrected in
eq.~(\ref{g3NC}).}~:
\ba
\int_0^1dxx^ng_1^{NC,+}(x,Q^2)&=&\sum_q\frac{((g_V^q)^
  2+(g_A^q)^2)a_n^{+q}}{4},~~~n=0,2~...\label{g1NC}\\
\int_0^1dxx^ng_2^{NC,+}(x,Q^2)&=&\sum_q\frac{((g_V^q)^2+(g_A^q)^2)
n(d_n^{+q}-a_n^{+q})}{4(n+1)}{\ },~~~n=2,4~...\label{g2NC}\\
\int_0^1dxx^ng_3^{NC,+}(x,Q^2)&=&
\sum_q \frac{g_V^qg_A^q(2a_{n+1}^{+q}+nd_{n+1}^{+q})}
{(n+2)}{\ },~~~n=0,2~...\label{g3NC}\\
\int_0^1dxx^ng_4^{NC,+}(x,Q^2)&=&
\sum_q g_V^qg_A^qa_{n+1}^{+q},~~~n=2,4~...
\label{g4NC}\\
\int_0^1dxx^ng_5^{NC,+}(x,Q^2)&=&\sum_q
\frac{g_V^qg_A^qa_n^{+q}}{2},~~~n=1,3~...
\label{g5NC}
\ea
and 
\ba
\int_0^1dxx^ng_1^{-}(x,Q^2)&=&\sum_q\frac{((g_V^q)^
  2+(g_A^q)^2)a_n^{-q}}{4},~~~n=1,3~...\label{g1-}\\
\int_0^1dxx^ng_2^{-}(x,Q^2)&=&\sum_q \frac{((g_V^q)^2+(g_A^q)^2)
n(d_n^{-q}-a_n^{-q})}{4(n+1)},~~~n=1,3~...\label{g2-}\\
\int_0^1dxx^ng_3^{-}(x,Q^2)&=&
\sum_q\frac{g_V^qg_A^q(2a_{n+1}^{-q}+nd_{n+1}^{-q})}
{(n+2)},~~~n= 1,3~...\label{g3-}\\
\int_0^1dxx^ng_4^{-}(x,Q^2)&=&
\sum_q g_V^qg_A^qa_{n+1}^{-q},~~~n= 1,3~...
\label{g4-}\\
\int_0^1dxx^ng_5^{-}(x,Q^2)
&=&\sum_q\frac{g_V^qg_A^qa_n^{-q}}{2},~~~n=0,2~...~.
\label{g5-}
\ea
Eqs.~(\ref{g3NC}, \ref{g3-}) differ from corresponding results obtained
in ref.~\cite{AR} for charged current interactions.
As evident from eqs.~(\ref{g1NC}--\ref{g5-}) only the structure functions
$g_2^i(x,Q^2)$ and $g_3^i(x,Q^2)$ contain twist 3 contributions.

Let us first consider the twist 2 contributions to the different
structure functions.
From eqs.~(\ref{g1NC},\ref{g2NC},\ref{g1-},\ref{g2-})
\be
\int_0^1dxx^ng_1^{i}(x,Q^2)=-\frac{n+1}{n}\int_0^1dxx^ng_2^{i}(x,Q^2),
\label{WW1}
\ee
follows,
where $ n=2,4... $ for $i = NC,+$  and $n=1,3...$ for $i = -$.
If an analytic continuation of the moment--index
$n$ to the complex plane is performed
for
eq.~(\ref{WW1}), one obtains the Wandzura--Wilczek relation~\cite{WW}
\be
g_2^i(x,Q^2)=-g_1^i(x,Q^2)+\int_x^1\frac{dy}{y}g_1^i(y,Q^2).
\label{WW}
\ee
Although it is formally consistent with the
Burkhardt-Cottingham sum rule \ct{BC}
\be
\label{e155}
\int_0^1dxg^i_2(x,Q^2)=0,
\ee
the $0th$ moment of the structure functions $g_2^i(x,Q^2)$ is not
described by the local
operator product expansion.

Eqs.~(\ref{g3NC}--\ref{g5NC}) and (\ref{g3-}--\ref{g5-}) 
yield the following
relations
between the
parity violating spin--dependent structure functions
\be
\int_0^1dxx^ng_4^{i}(x,Q^2) = 
\frac{n+2}{2}\int_0^1dxx^ng_3^{i}(x,Q^2),
\label{BK1}
\ee
with $ n=2,4... $ for the $i = NC, +$ exchange and $n = 1,3...$ 
for the $i = -$,
and
\be
\int_0^1dxx^ng_5^{i}(x,Q^2) = 
\frac{n+1}{4}\int_0^1dxx^{n-1}g_3^{i}(x,Q^2),
\label{BK2}
\ee
with $ n=1,3... $ for the $i = NC, +$  and $n = 2,4...$ for the $i = -$.
Analytic continuation leads to the new
relations
\be
\label{e300}
{g_3}^i(x,Q^2)=2x\int_x^1\frac{dy}{y^2}{g_4}^i(y,Q^2),
\ee
and
\be
{g_3}^i(x,Q^2)=4x\int_x^1\frac{dy}{y}{g_5}^i(y,Q^2).
\label{e31}
\ee
From \re{g3NC} and \re{g5NC} a new sum rule
\be
\int_0^1dx [g_3^{NC,+}(x,Q^2)
-2xg_5^{NC,+}(x,Q^2)]=0,
\label{e302}
\ee
is derived, which 
is not sensitive to the twist 3 contributions.

From eqs.~(\ref{g4NC}, \ref{g5NC},\ref{g4-}, \ref{g5-}) follows the
relation 
\be
\int_0^1dxx^ng_4^{i}(x,Q^2)=2\int_0^1dxx^{n+1}g_5^{i}(x,Q^2),
\label{dc1}
\ee
where $ n=2,4... $ for $i = NC, +$  and $n=1,3...$ for $i = -$.
Eq.~\re{dc1} yields
the  Dicus relation \ct{DIC}\footnote{This relation
corresponds to the Callan--Gross~\cite{CG}
relation for unpolarized structure functions since
the spin dependence enters the tensors of $g_4$ and $g_5$ in
$W_{\mu\nu}^{i}$,~eq.~(\ref{e6}), in terms of the overall
factor
$S.q$.}~:
\be
g_4^i(x,Q^2)=2xg_5^i(x,Q^2).
\label{DIC}
\ee
The latter equation is
a strict relation for the structure functions
$g_4$ and
$g_5$ in lowest order QCD,
since both structure functions
contain no
twist 3 contributions.

The  relations (\ref{WW}, \ref{e31}), and (\ref{DIC}) provide the
linear maps of a basis formed by the twist 2 contributions to
$g_1^i(x,Q^2)$ and $g_5^i(x,Q^2)$ to the complete set of the
twist 2 terms of the polarized structure functions in lowest order QCD.

Let us finally consider the twist~3
parts
of the structure functions
$g_2(x,Q^2)$ and $g_3(x,Q^2)$.
We define the twist 3  contribution to $g_2(x,Q^2)$ and $g_3(x,Q^2)$
by
\ba
g_2^{\rm III}(x,Q^2) &=& g_2(x,Q^2) + g_1(x,Q^2) - \int_x^1 \frac{dy}{y}
g_1(y,Q^2), \\
g_3^{\rm III}(x,Q^2) &=& g_3(x,Q^2)  - 4 x \int_x^1 \frac{dy}{y}
g_5(y,Q^2).
\label{T3g23}
\ea
The right hand sides of the eqs.~(\ref{T3g23}) describe the violation
of the Wandzura--Wilczek relation and the
twist 2 relation, eq.~(\ref{e31}),
in lowest order QCD, respectively, in the presence of twist~3 terms.

The twist~3 contribution to
charged current  and
electromagnetic structure functions
can be obtained using the relations
\ba
\int_0^1dx x^n(4g_5^-(x,Q^2)
-\frac{n+1}{x}g_3^-(x,Q^2))&=&\sum_q (n-1)d_n^{q-}, \nn\\
\int_0^1dx x^n(ng_1^\gamma(x,Q^2)
+(n+1)g_2^\gamma(x,Q^2))&=&\sum_q \frac{ne_q^2}{4}
d_n^{q+},~~~~n~=~2,4~...~.
\label{mm}
\ea
The definitions of  the matrix elements
$d_n^{q\pm}$ imply
\ba
\int_0^1dx x^n\{4g_5-\frac{n+1}{x}g_3\}^{\nu n-\nu p}=
\frac{12(n-1)}{n}
\int_0^1dx x^n\{ng_1+(n+1)g_2\}^{\gamma p-\gamma n},~~~~n~=~2,4~...~.
\label{mm1}
\ea

The analytic continuation in $n$ for this relation
finally yields the new integral relation among only twist~3 contributions
\be
g_3^{{\rm III}, \nu n - \nu p}(x,Q^2) =
12 \Biggl [ x g_2^{\rm III}(x,Q^2)
 - \int_x^1 dy
g_2^{\rm III}(y, Q^2) \Biggr ]^{\gamma p - \gamma n}.
\ee

\section{Sum Rules}
\label{sect6}

\vspace{1mm}
\noindent
A summary of sum rules derived by different techniques  in
the literature  is
given in Table~2. Here we  discuss first
the twist 2
contributions and consider deep-inelastic scattering off massless
quarks only. Most of the sum rules concern relations between the
different neutral and charged current structure functions. In some
cases relations to the first moment of a combination of polarized
parton densities as
\begin{eqnarray}
\label{eqGA}
g_A &=& \int_0^1 dx \left[ \Delta u(x) + \Delta \overline{u}(x)
                         - \Delta d(x) - \Delta \overline{d}(x) \right]
\nonumber\\
g_A^* &=& \int_0^1 dx \left[ \Delta u_V(x) - \Delta d_V(x) \right ]
\nonumber\\
g_A^8 &=& \int_0^1 dx \left[ \Delta u(x) + \Delta \overline{u}(x)
                         + \Delta d(x) + \Delta \overline{d}(x) 
                         - 2\left( \Delta s(x) + \Delta \overline{s}(x)
\right) \right]
\nonumber\\
g_A^{8*} &=& \int_0^1 dx \left[ \Delta u_V(x) + \Delta d_V(x) \right ]
\end{eqnarray}
are considered. 
In most of the studies three massless flavors were
assumed. Other sum rules being based on $SU(6)$ symmetry
were discussed
in refs.~\cite{a0,A1C}.

Before we will consider  specific sum rules, we summarize the basic
relations obtained in the previous sections. We consider the longitudinal
and transverse projection of the hadronic tensor, 
$W^{\parallel}_{\mu\nu}$ and $W^{\perp}_{\mu\nu}$, respectively.
The different relations between the twist~2 contributions to the
structure functions are illustrated schematically
in Figure~1.

\vspace{3mm}
\noindent
\renewcommand{\arraystretch}{2}
\[
\begin{array}{lclclcl}
 & &                        & & & \hspace{-8mm}
{\sf Dicus}  & \\
 & &    & &\hspace{20mm}   \swarrow  & &
  \searrow  \\
     W^{\parallel}_{\mu\nu}
&=& 
     i \varepsilon_{\mu\nu\alpha\beta} {\ds
\frac{q_{\alpha} P_{\beta}}{\nu} g_1(x)}
&+& {\ds \frac{P_{\mu} P_{\nu}}{\nu} g_4(x)}
&-&      g_{\mu\nu} g_5(x)  \\
 & &\hspace{20mm}  \uparrow  & &\hspace{12mm}   \uparrow  & & \\
 & &{\sf  Wandzura-Wilczek} & & {\sf this~~paper,~eq.~(\ref{e300})}& & \\
 & &\hspace{20mm}  \downarrow  & &\hspace{12mm}~\downarrow & & \\
{\ds W^{\perp}_{\mu\nu}} &=&{\ds
i \varepsilon_{\mu\nu\alpha\beta}
\frac{q_{\alpha}
S^{\perp}_{\beta}}{\nu} [g_1(x) + g_2(x)]} &+&
{\ds
\frac{P_{\mu} S^{\perp}_{\nu} + P_{\nu} S_{\mu}^{\perp}}{2 \nu}
g_3(x)}
& & \\
& & & & & & \\
 & & \hspace{1cm}
{\ds \overbrace{\Delta q~~~~+~~~~\Delta \overline{q}}}
     &|& &
 {\ds \overbrace{\Delta q~~~~-~~~~\Delta \overline{q}}} &
\end{array}
\]

\vspace{2mm}
\noindent
\begin{center}
{\sf Figure~1~: Relations between the twist~2 contributions of the
polarized structure functions}
\end{center}

\vspace{5mm}
\noindent

The structure functions $g_1(x),~g_4(x)$, and $g_5(x)$ contribute
to the longitudinal projection, whereas the transverse projection
contains $g_1(x)+g_2(x)$ and $g_3(x)$.  As shown above, the twist~2
contributions to the structure functions $g_1(x)$ and $g_2(x)$ are
$\propto \Delta q(x) + \Delta \overline{q}(x)$,
while $g_3(x), g_4(x)$ and $g_5(x)$
$\propto \Delta q(x) - \Delta \overline{q}(x)$. Except for valence
approximations, strict relations between the twist 2 contributions to the
structure functions can only be obtained either
for $g_1(x)$ and $g_2(x)$ 
{\it or} $g_3(x), g_4(x)$ and $g_5(x)$. While the two non-singlet
structure functions in $W^{\parallel}_{\mu\nu}$  are related
by a  factor\footnote{In the valence approximation a similar
relation is obtained for $g_1(x)+g_2(x)$ and $g_3(x)$.} only, the
two integral relations eqs.~(\ref{WW},\ref{e300}) 
relate the contributions
$\propto \Delta q(x) + \Delta \overline{q}(x)$ and
$\propto \Delta q(x) - \Delta \overline{q}(x)$
contained in $W^{\parallel}_{\mu\nu}$ and $W^{\perp}_{\mu\nu}$,
respectively.

A series of sum rules derived previously is  verified in the
context of the
operator product expansion, see Table~2. However, some relations
could not be confirmed. Among the latter are those which were derived
using the collinear parton model~\cite{a11}, in which the structure
functions cannot be  described correctly
for the case of
transverse nucleon polarization. We could also not confirm the
relations
\be
g_4(x,Q^2)-g_3(x,Q^2)=2xg_5(x,Q^2)
\ee
of ref.~\cite{a7}, which disagrees with Dicus' relation, and
\be
\int_0^1dxx^n\left[
\frac{n-7}{n+1}g_4(x,Q^2)+2g_3(x,Q^2)\right ]=0
\ee
from ref.~\cite{a77}. The latter two relations were derived using
the operator product expansion.

Some other relations, as the Burkhardt--Cottingham sum rule, for
different currents, and eq.~(23) in Table~2 are consistent with the
analytic continuations of the relations derived in section~5, but are
not obtainable in the framework of the local operator product expansion.

We would like to comment on  eqs.~(2) and (3) in Table~2 in more
detail.
They differ by a factor of two in the right hand side from the 
corresponding relations
(29) and (30),~Table~2,
\ba
24x[(g_1+g_2)^{ep}-(g_1+g_2)^{en}] &=&
 g_3^{\nu n}-g_3^{\nu p}, \label{ew22}
 \\
24x[g_2^{ep}-g_2^{en}] &=&
 (g_3-2g_4)^{\nu n}-(g_3-2g_4)^{\nu p}. \label{ew}
\ea
This difference has       a
subtle reason, which
can be traced back to the ansatz for the polarization
vector $n_\alpha$.
We will use the covariant parton model to illustrate this aspect.
As shown in ref.~\cite{BK} the results obtained in this approach are
completely
equivalent to those found using the local operator product
expansion for $m_q \rightarrow 0$ in lowest order QCD.

In refs.~\cite{DIC,WRAY,BAR} ( see also \cite{a1}) the
relation 
\be
n_\alpha \propto S_{\alpha}
\nn
\ee
was used,
with $ S_\alpha$  nucleon spin vector.
As a
result, linear relations between longitudinal and
transverse spin-dependent structure functions are implied,
which  violate
the Wandzura-Wilczek relation, the new relation (\ref{e300}), and yield
the extra factor of two in eqs.~(\ref{ew22},\ref{ew}).

Let us compare
 the derivation of
eq.~(\ref{ew22})
in
refs.~\cite{DIC,WRAY,BAR} with the calculation performed in the
covariant parton model in ref.~\cite{BK}.
The spin--dependent part of the
hadronic tensor has the following form
\ba
W_{\mu\nu}(q,P,S)=\sum_q\int d^4k\Delta f_q(q,k,S)w^q_{\mu\nu}
\delta[(k+q)^2-m_q^2],
\label{tt}
\ea
where $w^q_{\mu\nu}$ is quark tensor, and
$\Delta f_q(P,k,S)$ is
a function, which is related to the
polarized quark distributions.

In lowest order QCD the spin-dependent part
of the quark tensor  has the following form
\ba
{w_{\mu\nu}}^{q, spin}&=&\frac{1}{4}Tr[(1+\g
\frac{{\not \! n}}{m_q})
(\not \! {k} +m_q)\gamma_\mu(g_{V_1}+g_{A_1}\g)(\not \! {k}+\not \!
{q}+m_q)\gamma_\nu(g_{V_2}+g_{A_2}\g)]\nn\\
&=&
i\epb [2g_{A_1}g_{A_2}k_\alpha n_\beta+(g_{A_1}g_{A_2}+
g_{V_1}g_{V_2})q_\alpha n_\beta]\nn\\
&+& g_{V_1}g_{A_2}[2k_\mu n_\nu-(n\cdot q)g_{\mu\nu}]+
 g_{A_1}g_{V_2}[2n_\mu k_\nu-(n\cdot q)g_{\mu\nu}],
\ea
where  $n$ is the partonic spin vector.

While in the covariant parton model~\cite{a17} the (off--shell) parton
spin vector reads
\ba
n_\sigma=\frac{m_qp\cdot k}{\sqrt{(p\cdot k)^2 k^2-M^2k^4}}(k_\sigma-
\frac{k^2}{p\cdot k}p_\sigma),
\label{cov}
\ea 
the representation
\be
n_{\sigma} \approx \frac{m_q}{M} S_{\sigma}.
\ee
was used in~\cite{DIC,WRAY,BAR}.

One obtains
\ba
{W_{\mu\nu}}^{spin}
&=& \sum_q\frac{m_q}{2(Pq)M}
\int d^2k_\bot dy\Delta \tilde{f}_q(y,k^2)\Biggl \{
i\epb [\frac{g_{A_1}g_{A_2}}
{P.q}\left (
(S.q)q_\alpha P_\beta-(P.q)q_\alpha S_\beta \right)\nn\\
&+&
(g_{A_1}g_{A_2}+
g_{V_1}g_{V_2})q_\alpha S_\beta] +
g_{V_1}g_{A_2}[2xP_\mu S_\nu-(S.q)g_{\mu\nu}]\nn\\&+&
 g_{A_1}g_{V_2}[2xS_\mu P_\nu-(S.q)g_{\mu\nu}]\Biggr \}.
\label{naive1}
\ea
$\Delta \tilde{f}_q(y, k^2)$ is independent from
the nucleon spin, with
$k=xP + yq'  + k_\bot$ and $q'=q+xP$. 

The projections on the structure function
combinations for photon and charged current interactions are
\ba
{W_{\mu\nu}}^{\gamma,~spin}
&=& i\epb
q_\alpha S_\beta
\sum_q\frac{e_q^2m_q}{2(Pq)M}
\int d^2k_\bot dy\Delta \tilde{f}_q(y,k^2)
\label{naiveem}
\ea
and
\ba
{W_{\mu\nu}}^{W^{\pm},~spin}
&=& 
[\frac{i\epb }{2}(\frac{S.q}{P.q}q_\alpha P_\beta +q_\alpha S_\beta)-
2x\frac{P_\mu S_\nu +P_\nu S_\mu}{2}\nn\\
&+&g_{\mu\nu}(S.q)]
\sum_q\frac{m_q}{(P.q)M}
\int d^2k_\bot dy\Delta \tilde{f}_q(y,k^2).
\label{naivech}
\ea 
From (\ref{naiveem}) and (\ref{naivech})
\ba
&12x[(g_1+g_2)^{ep}-(g_1+g_2)^{en}] =  g_3^{\nu n}-g_3^{\nu p}
\label{naiver}
\ea
follows,
which differs from eq.~(\ref{ew22}).

In the covariant parton model, on the other hand, the distribution
\ba
\Delta f(p\cdot k,k\cdot S,k^2)=-\frac{(n\cdot S)}{M^2}
\tilde{f}^{\rm cov}(p\cdot k,k^2),
\ea
depends on the nucleon spin. As lined out in refs.~\cite{a17} 
and \cite{BK} the structure functions $g_i(x)$ are represented
by
\ba
g_1(x)&=&\frac{a\pi xM^2}{8}\int_x^1dy(2x-y)\tilde{h}(y), \nn\\
g_2(x)&=&\frac{a\pi xM^2}{8}\int_x^1dy(2y-3x)\tilde{h}(y), \nn\\
g_1(x)+g_2(x)&=&\frac{a\pi xM^2}{8}\int_x^1dy(y-x)\tilde{h}(y), \nn\\
g_3(x)&=&\frac{b\pi x^2M^2}{2}\int_x^1dy(y-x)\tilde{h}(y), \nn\\
g_4(x)&=&\frac{b\pi x^2M^2}{4}\int_x^1dy(2x-y)\tilde{h}(y), \nn\\
g_5(x)&=&\frac{b\pi xM^2}{8}\int_x^1dy(2x-y)\tilde{h}(y) ,
\label{q6}
\ea
where $a=g_{A_1}g_{A_2}+g_{V_1}g_{V_2}$,
$b=g_{V_1}g_{A_2}+g_{V_2}g_{A_1}$,
$ y=x+k^2_\bot/(xM^2)$ and 
\be
\tilde{h}(y)=
\int dk^2\tilde{\Delta f^{\rm cov}}(y,k^2).
\ee
From these relations eq.~(\ref{ew22})
is obtained. This example  clearly demonstrates,
how carefully calculations
in partonic approaches have to be performed to obtain correct results
for the polarized
structure functions contributing to $W_{\mu \nu}^{\perp}$.

\newpage
\renewcommand{\arraystretch}{1.5}

\vspace{3mm}
\noindent
{\sf Table~2~: A comparison of
different structure function relations (twist 2) derived in the
literature with the results obtained in the local operator product
expansion in
section~5. The signs in the last column
mark agreement or disagreement.}

\vspace{5mm}
\begin{center}
\begin{tabular}{||r|c||c|c||}
\hline \hline
 &
{\sf sum rule } & {\sf ref.} & $m_q=0$  \\
\hline \hline
 & & &   \\
1 & $ g_4=2xg_5$      &  {\sf \cite{DIC,a3,a77,FRANKF}}
& $+$    \\
 & & &   \\
2 & $12x[(g_1+g_2)^{ep}-(g_1+g_2)^{en}] =  g_3^{\nu n}-g_3^{\nu p}$ &
{\sf \cite{DIC,WRAY,BAR}}    & $-$ \\
 & & &   \\
3 &$12x[g_2^{ep}-g_2^{en}] = (g_3-2g_4)^{\nu n}-(g_3-2g_4)^{\nu p}$ &
 & $-$     \\
& & &   \\
\hline
 & & &   \\
4 &$\ds 12x(g_1^{ep}-g_1^{en}) = g_4^{\nu n}-g_4^{\nu p}$ &
{\sf \cite{WRAY}}
& $+ $      \\
 & & &   \\
 5 &$\ds 3 \int_0^1 dx (g_1^{ep} - g_1^{en})
- \int_0^1 dx (g_1^{\nu p} + g_1^{\nu n})= - \frac{1}{6}g_A^8 $ & &
$- $      \\
 & & &   \\
 6 &$\ds 6\int_0^1 dx (g_2^{ep}-g_2^{en})-
\int_0^1dx(g_1^{\nu p}-g_1^{\nu n})=-g_A^*$      &  &  $+ $  \\
 & & &   \\
7 & $\ds 12\int_0^1dx(g_2^{ep}-g_2^{en})-
\int_0^1\frac{dx}{x}(g_4^{\nu p}-g_4^{\nu n})=-2g_A$      &  &
$+ $  \\
 & & &   \\
8 & $12x[g_1^{ep}-g_1^{en}] =  g_3^{\nu n}-g_3^{\nu p}$ &
    & $-$    \\
\hline
 & & &   \\
9 & $ \ds
\int_0^1dx [(g_1+g_2)^{\bar\nu p}-(g_1+g_2)^{\nu p}]
=g_A^* $ & 
{\sf \cite{a1B}}
& $+$
   \\
 & & &   \\
 10 &$  \ds \int_0^1\frac{dx}{x} [g_3^{\bar\nu p}+g_3^{\nu p}]=
- g_A^{8*}$
&   &
$+$ \\
 & & &   \\
\hline
 & & &   \\
11 & $\ds \int_0^1 dx g_2^{\gamma}=0$ &{\sf \cite{BC} }
 & $ $  \\
 & & 
 &   \\
\hline \hline
\end{tabular}
\end{center}

\newpage
\begin{center}
\begin{tabular}{||r|c||c|c||}
\hline \hline
&
{\sf sum rule } & {\sf ref.} & $m_q=0$  \\
\hline \hline
 & & &   \\
12 & $ \ds \int_0^1dxx(g_1+2g_2)^{\nu p-\bar \nu p}=0$  & {\sf \cite{A1C}}
  & $+$
\\
 & &
 &   \\
13& $ \ds \int_0^1dx(g_3-2xg_5)^{\nu p+\bar \nu p}=0$   &  & $+$
\\
 & & &  \\
14& $\ds \int_0^1d x ( g_4-g_3)^{\nu p + \bar \nu p} = 0 $ &
   & $+$     \\
 & & &   \\
\hline
  & & & \\
15 &$\ds \int_0^1dx(g_5^{\nu p}-g_5^{\nu n})=g_A$ &
{\sf \cite{BAR}}
   & $+$     \\
 & & &  \\
16 &$\ds \int_0^1dx\frac{[(  g_4-g_3)^{\nu p}-(  g_4-g_3)^{\nu n}]}{x}=0$
&  & $-$   \\
 & & &  \\
17 & $\ds \int_0^1dx\frac{(g_3^{\nu p }-g_3^{\nu n})}{x}=2g_A$ & &  $-$
\\
  & & & \\
\hline
  & & & \\
 18 &$\ds g_4-g_3=2xg_5$  & {\sf \cite{a7}       } &$-$ \\
 & & &  \\
\hline
 & & &  \\
19 &$ \ds \int_0^1dxx^n(\frac{n-7}{n+1}g_4+2g_3)=0$  &
{\sf \cite{a77}}
 & $-$  \\
 & & &  \\
\hline
 & & &  \\
20 & $g_3=2xg_5$      & {\sf \cite{a1}} & $-$     \\
 & & &  \\
21 &  $g_3=g_4$ & & $-$        \\
 & & &  \\
22 & $g_2^{\gamma }=g_2^{\gamma Z}=0$ & &  $-$      \\
 & & &  \\
23 & $g_1^{W^\pm}=-2g_2^{W^\pm}$ & & $-$  \\
 & &  & \\
\hline \hline
\end{tabular}
\end{center}

\newpage
\begin{center}
\begin{tabular}{||r|c||c|c||}
\hline \hline
&
{\sf sum rule } & {\sf ref.} & $m_q=0$  \\
\hline \hline
 & & &   \\
24 &$ \ds \int_0^1dx(g_3-g_4)^{(\nu+\bar\nu),\gamma,Z}=0$
& {\sf \cite{FRANKF}}
 &         \\
 & & &   \\
25 &$ \ds \int_0^1dxg_2^{\nu+\bar\nu }=0$ & &         \\
   & & & \\
26 & $ \ds \int_0^1 dx x \left [g_1 + 2 g_2
\right ]^{W^--W^+} = 0$& 
 &$+$ \\
 & & &   \\

\hline
 & & &   \\
27
 & $ \ds \int_0^1dx(g_3-2xg_5)^{(\nu+\bar\nu),\gamma,Z}=0$&
{\sf           } & $+$  \\
 & & &   \\
28 & $ \ds \int_0^1 dx \frac{g_3^{\nu p} - g_3^{\nu n}}{x} = 4 g_A$
& {\sf  this paper}
           & $+$  \\
 & & &   \\
29 & $24x[(g_1+g_2)^{ep}-(g_1+g_2)^{en}] =  g_3^{\nu n}-g_3^{\nu p}$ &
    & $+$     \\
 & & &   \\
30 &$24x[g_2^{ep}-g_2^{en}] = (g_3-2g_4)^{\nu n}-(g_3-2g_4)^{\nu p}$ &
 & $+$     \\
& & &   \\
31 &$\ds  \int_0^1 dx (g_1^{ep} + g_1^{en})
- \frac{2}{9}\int_0^1 dx (g_1^{\nu p} + g_1^{\nu n})= 
\frac{1}{18}g_A^8 $ & &
$+$          \\
 & & &   \\
\hline\hline
\end{tabular}
\end{center}

\vspace{12mm}
Recently, a sum rule for the valence part of the structure functions
$g_1(x,Q^2)$ and $g_2(x,Q^2)$
\be
\int_0^1dx x(g_1^V(x)+2g_2^V(x))=0
\label{etl}
\ee
was discussed in ref.~\ct{ELT}\footnote{
This sum rule was found  firstly in ref.~\ct{A1C} for a
specific flavor combination.}
in the context of a `field-theoretic' framework.

Since the valence parts
$g_1^V(x)$ and $g_2^V(x)$ cannot be isolated for electromagnetic
interactions from the complete structure functions, a formulation
of eq.~(\ref{etl}) with the help of the local operator product expansion
is thus
not straightforward. On the other hand, one
may consider
eqs.~(\ref{g1-}) and (\ref{g2-}) from which
\be 
\int_0^1dxx^n(g_1^-(x,Q^2)+2g_2^-(x,Q^2))=\sum_q \frac{((g_V^q)^2+(g_A^q)^2)
(nd_n^{-q}-(n-1)a_n^{-q})}{4(n+1)},~~~~n=1,3~...
\label{etl1}
\ee
follows for the charged current case.
It is easily seen that the left hand side of eq.~(\ref{etl1})
includes only valence  quark contributions. We may even rewrite
eq.~\re{etl1}  for individual
quark flavors separately
\be 
\int_0^1dxx^n(g_1^{Vq}(x,Q^2)+2g_2^{Vq}(x,Q^2))=
\frac{e_q^2(nd_n^{Vq}-(n-1)a_n^{Vq})}{4(n+1)},~~~~ n = 1,3 ...~,
\label{etl2}
\ee
where $d_n^{Vq}, a_n^{Vq} $ are the valence parts of the 
corresponding matrix elements.

For the first  moment of this equation one obtains
\be 
\int_0^1dxx(g_1^{Vq}(x,Q^2)+2g_2^{Vq}(x,Q^2))=\frac{e_q^2}{8}
d_1^{Vq}.
\label{etl3}
\ee
The right hand
side of eq.~\re{etl3} vanishes in the case of massless quarks, because
\ba
\langle PS|\bar q (\gamma_\beta\gamma_5 D^\mu
-\gamma_\mu\gamma_5 D^\beta )q|PS \rangle
=      d_1^{Vq}(S^\beta P^\mu-
S^\mu P^\beta)
= m_q
\langle PS|\bar q i\gamma_5\sigma^{\beta\mu}q|PS \rangle.
\label{em}
\ea
Therefore eq.~\re{etl} can be derived in the operator product expansion.

We mention that the  local operator product expansion for the valence 
parts of the $g_1$ and $g_2$
contains only the {\it odd} moments of the
structure functions, whereas the
operator product expansion for the complete structure functions
$g_1(x,Q^2)$ and $g_2(x,Q^2)$  \re{g1NC}, \re{g2NC} concerns the
{\it even} moments.

The  matrix element in \re{em} is related to the
first moment
of the  structure function $h_1(x)$~\ct{JAFFE}
\be
\langle
PS|\bar q i\gamma_5\sigma^{\beta\mu}q|PS \rangle = \frac{\delta q}{M}
(P^\beta S^\mu-P^\mu S^\beta),
\label{mee}
\ee
where
\be
\delta q=\int_0^1dx(h_1^q(x)-\bar h_1^q(x)),
\label{de}
\ee
and
$h_1^q(x)$, $\bar h_1^q(x)$ are the
quark and antiquark transversity functions,
respectively, which can be measured in the
Drell-Yan process.

\section{Conclusions}
\label{sect7}

\vspace{1mm}
\noindent
We have derived the twist 2 and twist 3 contributions to the polarized
structure functions in lowest order QCD for the general case of
both neutral and charged current
electroweak interactions. The results were obtained using the local
operator product expansion. The twist 2 parts of the five structure
functions per current combination are not independent but related by
three linear operators, the Wandzura--Wilczek relation, a relation
by Dicus and a new integral relation. A new relation between
twist 3 contributions to the structure function $g_2(x,Q^2)$
and $g_3(x,Q^2)$ was derived. It was shown, that a relation
between the valence contributions of the structure functions
$g_1(x,Q^2)$ and $g_2(x,Q^2)$, eq.~(\ref{etl}), can be derived
using the local operator product expansion.

\vspace{4mm}
\noindent
{\bf Acknowledgements.}~We would like to thank Prof. Paul S\"oding 
for his constant support of the present project. One of us (N.K.) would 
like to thank DESY-Zeuthen for the hospitality extended to him.

\end{document}